\documentclass[aps,twocolumn,groupedaddress,floatfix]{revtex4-1}
\usepackage{amssymb,amsmath}
\usepackage{graphicx}
\usepackage{subfigure}
\usepackage[english]{babel} 
\usepackage{float}
\usepackage{color}

\usepackage{lipsum}
\begin{document}
%\DeclareRobustCommand{\baselinestretch{2}}
\newcommand{\be}{\begin{equation}}
\newcommand{\ee}{\end{equation}}
\newcommand{\rojo}[1]{\textcolor{red}{#1}}

\title{Exploring the interplay between fractionality and $\cal{PT}$ symmetry in magnetic metamaterials}

\author{Mario I. Molina}
\affiliation{Departamento de F\'{\i}sica, Facultad de Ciencias, Universidad de Chile, Casilla 653, Santiago, Chile}

\date{\today }

\begin{abstract}
We study a nonlinear magnetic metamaterial modeled as a split-ring resonator array, where the standard discrete laplacian is replaced by its fractional form. We find a closed-form expression for the dispersion relation as a function of the fractional exponent $s$ and the gain/loss parameter $\gamma$ and examine the conditions under which stable magneto-inductive waves exist. The density of states is computed in closed form and suggests that the main effect of fractionality is the flattening of the bands, while gain/loss increase tends to reduce the bandgaps.  The spatial extent of the modes for a finite array is computed by means of the participation ratio $R$, which is also obtained in closed form. For a fixed fractionality exponent, an increase in gain/loss $\gamma$ decreases the overall $R$, from the number of sites $N$ towards $N/2$ at large $\gamma$. The nonlinear dynamics of the average magnetic energy on an initial ring during a cycle shows a monotonic increase with $\gamma$, and it is qualitatively similar for all fractional exponents. This is explained as mainly due to the interplay of nonlinearity and ${\cal PT}$ symmetry.

\end{abstract}

\maketitle

{\em Introduction}. 
More than $300$ years have elapsed since the first discussions among mathematicians concerning the possible extension of Newton's calculus to non-integer derivatives, for instance, asking ``What would a half derivative of a function be like?''. Initial studies on simple cases suggested the feasibility of this idea. Since then, rigorous studies have been conducted by mathematicians like Laplace, Euler, Riemann, and Caputo, to name a few, and have promoted fractional calculus from a curious mathematical object to a whole research field. Although classical physical systems are usually described by means of differential equations of integer order, the relevance of fractional order is found in its capacity to address complex nonlinear systems with memory effects that are hard to treat with conventional methods. Applications have been found in fluids mechanics\cite{fluid2}, fractional kinetics and anomalous diffusion\cite{metzler,sokolov,zaslavsky}, strange kinetics\cite{shlesinger}, fractional quantum mechanics\cite{laskin,laskin2}, Levy processes in quantum mechanics\cite{levy}, plasmas\cite{plasmas}, electrical conduction in cardiac tissue\cite{cardiac}, and biological invasions\cite{invasion}, among others.

Another relatively recent advance is the realization that a quantum Hamiltonian can be non-hermitian and yet possess real eigenvalues. It was shown that if a system is invariant under the combined operations of spatial reflection and time reversal symmetries, its eigenvalue spectrum is always real, provided there is no spontaneous symmetry breaking. The ${\cal PT}$ symmetry conditions translate into the requirement that the real part of the complex potential be an even function in space while its imaginary part is odd in space. In an optics context, this means an index of refraction whose real(imaginary) part is even (odd) in space. Under these conditions, the balance between gains and losses leads to stable dynamics. 
Currently, numerous PT-symmetric systems have been explored in several settings, from electronic circuits\cite{circuits}, optics\cite{optics1,optics2,optics3,optics4,optics5}, magnetic metamaterials\cite{MM}, to solid-state and atomic physics\cite{solid1,solid2}, among others. 

\begin{figure}[t]
 \includegraphics[scale=0.14]{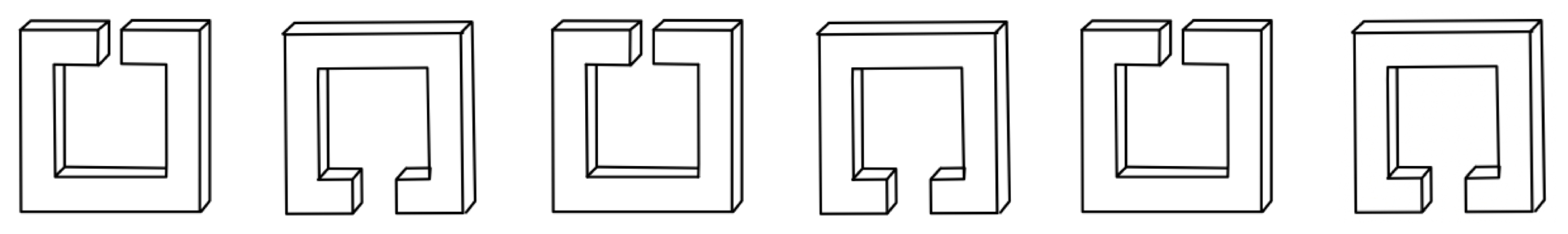}\\
  \includegraphics[scale=0.14]{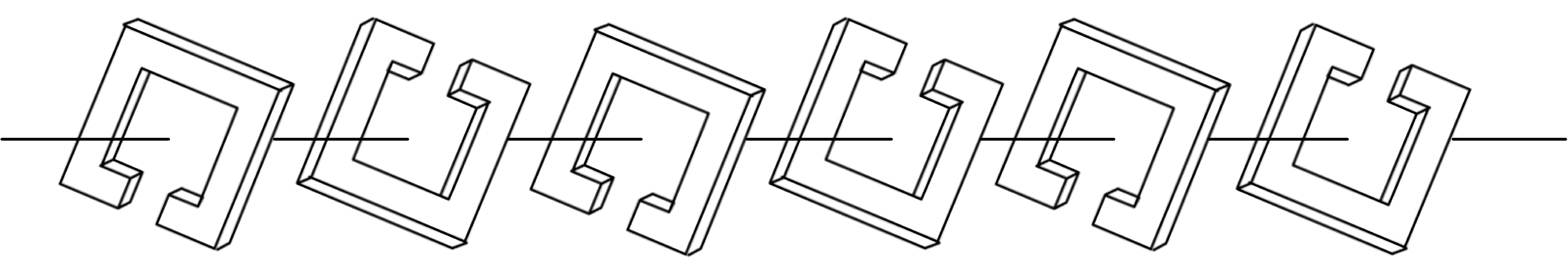}
  \caption{One-dimensional split-ring resonator arrays. Top: All SRRs lying on a common plane ($\lambda<0$). Bottom:   All SRRs are parallel but centered around a common axis ($\lambda>0$).  
  }
  \label{fig1}
\end{figure}

Metamaterials constitute a new class of man-made materials that are characterized for having interesting thermal, optical, and transport properties that make them attractive candidates for current and future technologies. Among them, we have 
magnetic metamaterials (MMs) whose magnetic response can be tailored to a certain extent. A simple realization of such a system consists of an array of metallic split-ring resonators (SRRs) coupled inductively\cite{SRR1, SRR2, SRR3}. For instance, this MM can feature negative magnetic response in some frequency window, making them attractive for use as a constituent in negative refraction indexmaterials\cite{negative_refraction}. In order for SRRs to be practical, one must overcome the problem of ohmic and radiative loss. A possible solution that has been considered is to endow the SRRs with external gain, such as tunnel (Esaki) diodes\cite{losses1,losses2} to compensate for such losses.  The simplest MM model uses an array of split-ring resonators (Fig.\ref{fig1}), with each resonator consisting of a small, conducting ring with a slit. Each SRR unit in the array can be mapped to a resistor-inductor-capacitor (RLC) circuit featuring self-inductance $L$, ohmic resistance $R$, and capacitance $C$ built across the slit. 
In the absence of resistance, each unit will possess a resonant frequency $\omega_{0}\sim 1/\sqrt{L C}$. We insert a nonlinear dielectric (Kerr) inside the slits with permittivity $\epsilon=\epsilon_{l} + a |{\bf E}|^2$, where ${\bf E}$ is the electric field inside the slit\cite{shadrivov}, and $\epsilon_{l}$ is the linear permittivity. 

{\em The model.}\ To keep things simple, we assume that the magnetic component of the incident electromagnetic wave is perpendicular to the SRRs plane. In that case, ${\bf E}$ originates from the oscillations of the magnetic field only and is parallel to the slit.
In addition, the configurations shown in Fig.\ref{fig1} with alternating orientations of the slits serves to decrease electrical dipole-dipole effects. Under these conditions, and in the absence of fractional effects, 
the charge $q_{n}$ residing at the $nth$ ring can be written in dimensionless form as\cite{gts_2006}
\be
{d^2\over{d t^2}}\left( q_{n} + \lambda (q_{n+1} + q_{n-1})\right) + q_{n}+(-1)^n \gamma {d q_{n}\over{dt}}+ \chi\ q_{n}^3 = 0 \label{eq1}
\ee
where $q_{n}$ is the dimensionless charge of the nth ring,  $\lambda$ is the coupling between neighboring rings which originates from the dipole-dipole interaction, $\gamma$ is the gain/loss parameter, and $\chi$ is the nonlinear parameter that originates from the Kerr nonlinearity of the permittivity of the medium inside the slits. Even though model (\ref{eq1}) is quite simple, it still retains the main physics, making it possible to extract analytical results.
%%%%%%%%%%%%%%%%%%%%%%%%%%%%%%%%%
%%%%%%%%%%%%%%%%%%%%%%%%%%%%%%%%

The term $\lambda (q_{n+1} + q_{n-1})$ in Eq.(\ref{eq1}) is essentially a discrete laplacian $\Delta_{n} q_{n} = q_{n+1}-2 q_{n}+q_{n-1}$. Thus, we can rewrite Eq.(\ref{eq1}) as
\be
{d^2\over{d t^2}}\left( q_{n} + 2 \lambda q_{n} + \lambda \Delta_{n} q_{n}  \right) + q_{n} +
(-1)^n \gamma {d q_{n}\over{dt}}+ \chi q_{n}^3 = 0 \label{eq2}
\ee
Let us now introduce fractionality. 
We proceed to replace $\Delta_{n}$ by its fractional form $(\Delta_{n})^s$ in Eq.(\ref{eq2}). The discrete fractional laplacian is given by\cite{discrete laplacian}
\be
(-\Delta_{n})^s q_{n}=\sum_{m\neq n} K^s(n-m) (q_{n}-q_{m}),\hspace{0.5cm}0<s<1 \label{delta}
\ee
where,
\be
K^{s}(m) = {4^{s} \Gamma((1/2)+s)\over{\sqrt{\pi}|\Gamma(-s)|}}\  {\Gamma(|m|-s)\over{\Gamma(|m|+1+s)}}.\label{K}
\ee

Equation(\ref{eq1}) becomes
\begin{eqnarray}
{d^2 \over{d t^2}} (\ q_{n} + 2 \lambda q_{n}+\lambda \sum_{m\neq n}K^s(m-n) (q_{m}-q_{n})\ ) + q_{n}& &\nonumber\\
 +(-1)^n \gamma {d q_{n}\over{dt}}+\ \chi\ q_{n}^3 = 0.\ \ \ \ \ \ \ \ & &\label{eq7}
\end{eqnarray}

Note that the presence of the kernel $K^{s}(m)$
 introduces nonlocal coupling effects. At long distances, $K^{s}(m)\approx 1/|m|^{2 s+1}$. Thus, at $s=1$ (non-fractional case), we have the typical magnetic dipole-dipole coupling. In contrast, at the other extreme ($s\approx 0$), we have a very long coupling range decrease of $1/|m|$, coupling virtually all sites with nearly similar values.   
 
%%%%%%%%%%%%%%%%%%%%%%%%%%%%%%%%%%%%%%
\begin{figure}[t]
 \includegraphics[scale=0.25]{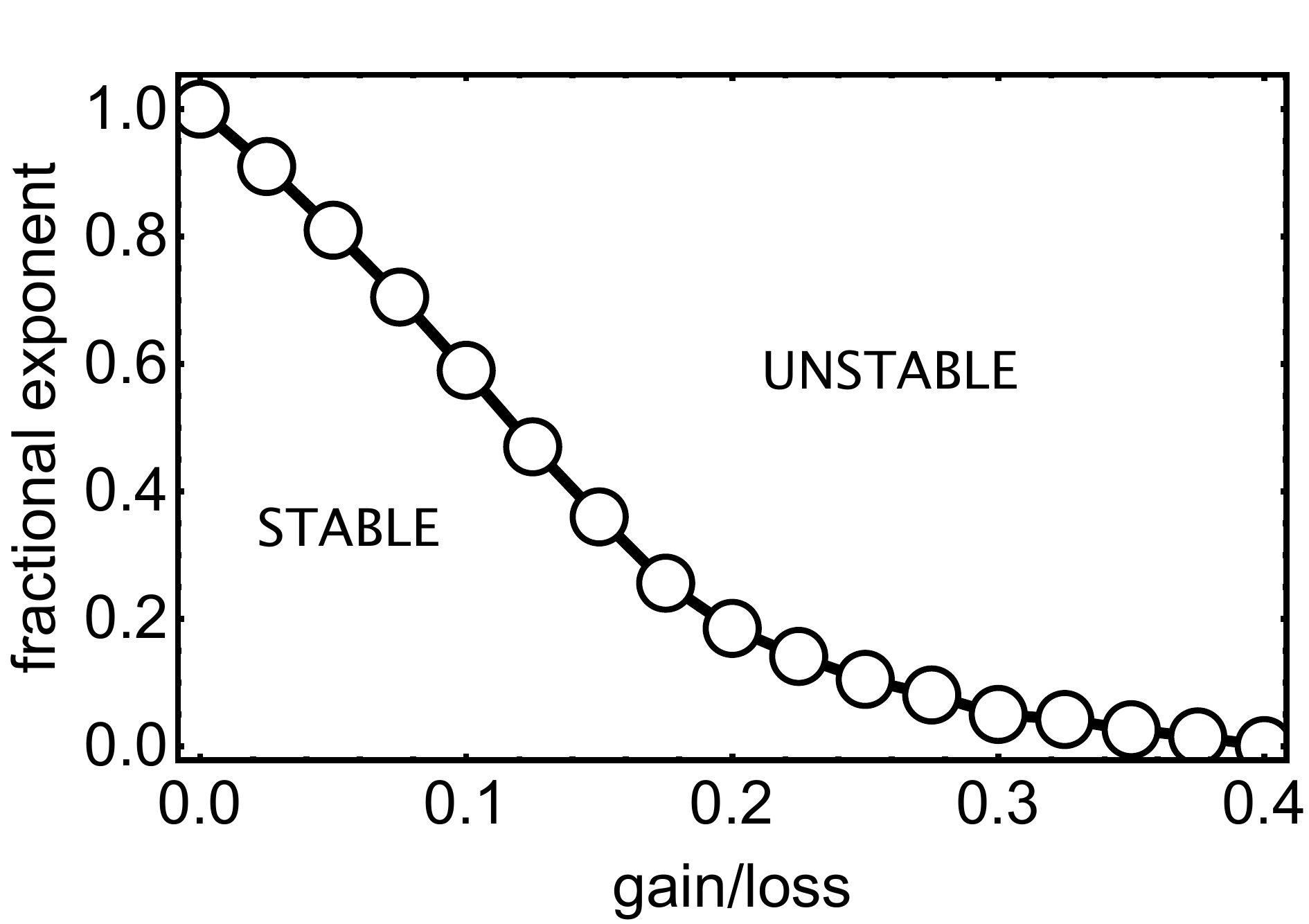}
  \caption{Stability diagram in $\gamma$-$s$ phase space, showing the regions in parameter space where the eigenmodes are real or complex. }
  \label{fig2}
\end{figure}
%%%%%%%%%%%%%%%%%%%%%%%%%%%%%%%%%%%%%%%%%%%%%%%%

Let us first look for linear waves ($\chi=0$). Given the binary distribution of gain/loss, we pose a solution of the form $q_{n} = A \cos(k n-\Omega t)$ for $n$ even, and $q_{n} = B \cos(k n-\Omega t)$ for $n$ odd. This leads to the stationary equations
\begin{eqnarray}
& &(-\Omega^2+ i \gamma \Omega +1) A +\nonumber\\
& & + B (-\lambda \Omega^2\sum_{q}K^{s}(q)(e^{i k q}-1)
-2 \lambda \Omega^2=0
\end{eqnarray}
\begin{eqnarray}
& & (-\Omega^2- i \gamma \Omega +1) B + \nonumber\\
& & +A (-\lambda \Omega^2\sum_{q}K^{s}(q)(e^{i k q}-1)-2 \lambda \Omega^2=0
\end{eqnarray}
After imposing that the determinant of this system be equal to zero, one obtains the dispersion relation in closed form:
\be
\Omega^2={2\over{2-\gamma^2\pm \sqrt{\gamma^4-4 \gamma^2 \lambda^2 (1-2 H(k))^2}}}\label{om2}
\ee
where $H(k)=\sum_{m\neq 0} K^{s}(m) \sin^{2}(k\ m/2)$. $H(k)$ can be expressed in closed form in terms of regularized hypergeometric functions:
\begin{eqnarray}
\lefteqn{H(k) = (1/2) {\Gamma(2 s)\over{(\Gamma+1)\Gamma(s)}}}\nonumber\\
&\times&  [ 1-s ( e^{-i k} {_{2}}F_{1} (1,1-s,s+2; e^{-i k})+\nonumber\\
& & e^{i k} {_{2}}F_{1} (1,1-s,s+2; e^{i k}) )].\label{H(k)}
\end{eqnarray}
Stable modes occur for real $\Omega$. In Fig. 2 we show a stability diagram in gain/loss- fractional exponent phase space. For $\gamma=0$, the modes are stable for all $s$ values. As the gain/loss parameter increases from zero, the fractional exponent must decrease from $s=1$ to maintain the system stability.

{\em Density of states}.\ The density of states $\delta(\Omega^2)=(1/N) \sum_{n} \delta(\Omega^2-\Omega^2(k)$ is complex in general since $\Omega^2(k)$ is complex. We define partial densities of states for the real and imaginary part of the spectrum:
\begin{eqnarray}
\delta_{R}(\Omega^2) &=& (1/N) \sum_{k} \delta(\Omega^2-Re[\Omega^2(k)])\nonumber\\
\delta_{I}(\Omega^2) &=& (1/N) \sum_{k} \delta(\Omega^2-Im[\Omega^2(k)]).\label{DOS}
\end{eqnarray}
By using the analytical expressions (\ref{om2}),(\ref{H(k)}) for $\Omega^2({\bf k})$ into Eq.(\ref{DOS}),
we compute numerically $\delta_{R}(\Omega^2)$ and $\delta_{I}(\Omega^2)$ for a fixed gain/loss value $\gamma=1$, and several fractional exponents $s$, ranging from $s\approx 1$ (standard case) down to $s\approx 0$. Results are shown in Fig.3. In the absence of gain/loss, $\delta^{R}(\Omega^2)$ splits into two bands which become flatbands at $s\approx 0$, with values $\Omega^2=1/(1+2 \lambda)$ and $\Omega^2=1/(1-2\lambda)$. If we now increase the gain/loss parameter, the flatbands start to converge towards each other, merging completely at $\gamma=2(1-\sqrt{1-4 \lambda^2})$, with $\Omega^2=1/\sqrt{1-4 \lambda^2}$. Further increase of $\gamma$ shifts the position of this flat band towards the low-frequency sector. While fractionality tends to flatten the bands, gain/loss reduces the separation between the bands.

{\em Participation Ratio}. The spatial extent of the $\alpha$  mode can be computed by means of the participation ratio $R$, defined as
\be
R^{\alpha} = {(\sum_{n} |q^{\alpha}_{n}|^2)^2\over{\sum_{n} |q^{\alpha}_{n}|^4}}.
\ee
For a completely localized mode, $R=1$, while for complete delocalization, $R=N$, the total number of sites. for definiteness, let us take $N=233, \lambda=0.2$ and compute $R$ vs. mode frequency index for three fixed values of $s$ and several $\gamma$ values for each of them. Results are shown in Fig.4. Generally, we see that $R$ decreases with an increase in $\gamma$. An interesting feature is that $R$ is contained between $N$ and $N/2$ as the gain/loss parameter increases. This shows a degree of localizing effect of gain/loss for all fractional exponents.
%%%%%%%%%%%%%%%%%%%%%%%%%%%%%%%%%%%%%%
\begin{figure}[t]
 \includegraphics[scale=0.15]{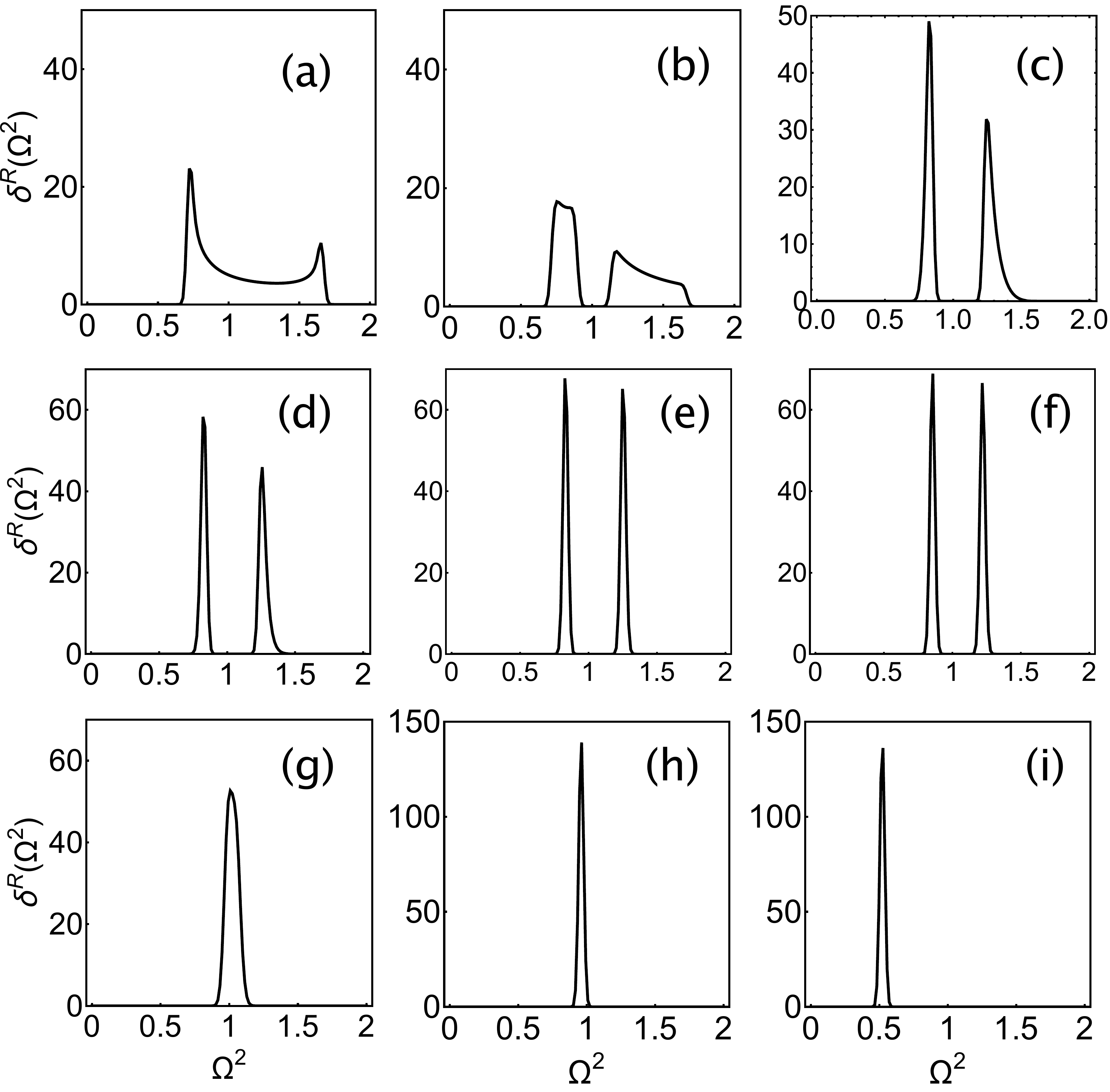}
  \caption{Real part of the density of states, for several fractional exponents and several gain/loss parameters:
  $(s,\gamma)$=$(1,0)(a), (0.5,0)(b),(0.1,0)(c),(0.05,0)(d),(0.01,0)(e),\newline
 (0.01,0.01)(f),(0.01,0.02)(g),(0.01,0.04)(h),(0.01,1)(i). (\lambda=0.2)$}
  \label{fig3}
\end{figure}
%%%%%%%%%%%%%%%%%%%%%%%%%%%%%%%%%%%%%%%%%%%%%%%%
%%%%%%%%%%%%%%%%%%%%%%%%%%%%%%%%%%%%%%
\begin{figure}[t]
 \includegraphics[scale=0.122]{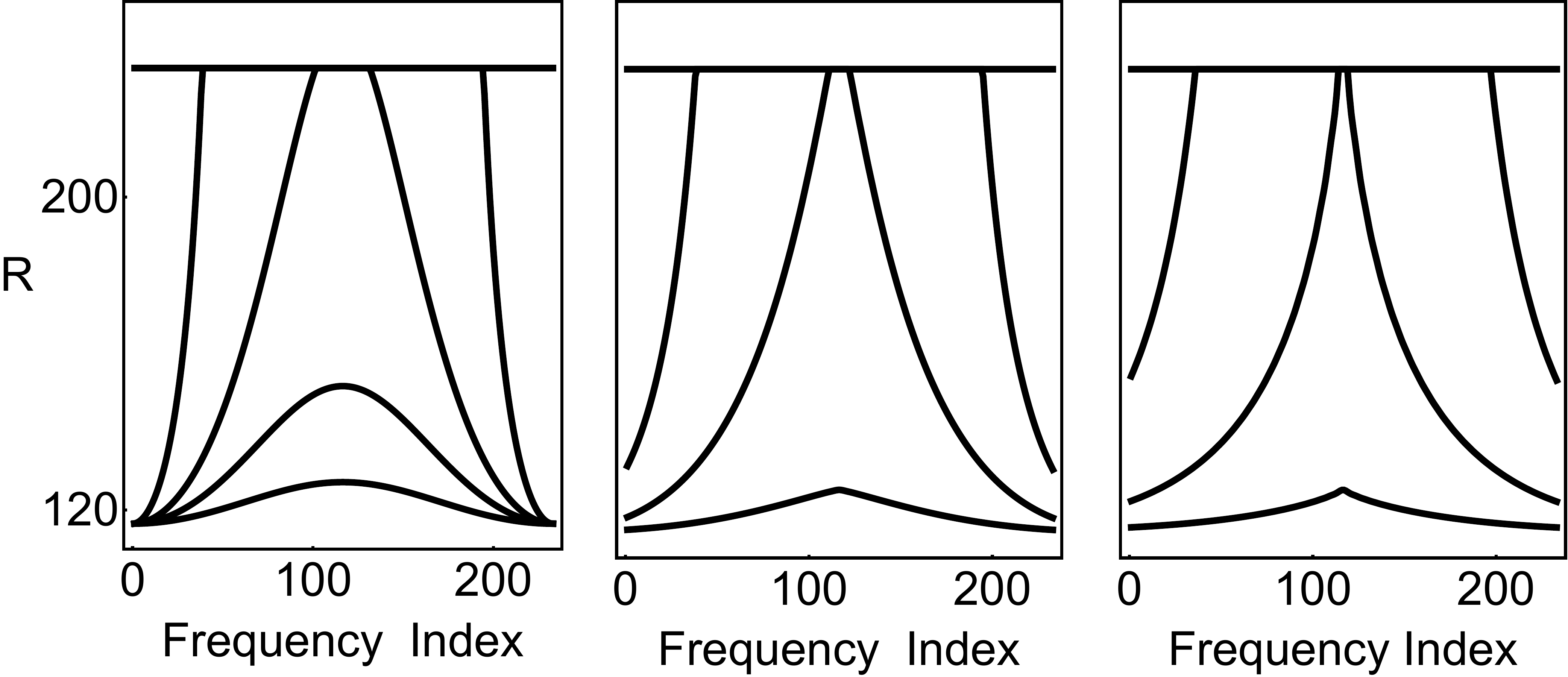}
  \caption{Participation Ratio: $s=1$ (left), $s=0.6$ (middle), $s=0.4$ (right), and $\gamma=0\ (\text{upper curve}),0.2,0.4,0.6,1.0\ (\text{lowest curve})$ ($\lambda=0.2, N=233$).}
  \label{fig4}
\end{figure}
%%%%%%%%%%%%%%%%%%%%%%%%%%%%%%%%%%%%%%%%%%%%%%%%

{\em Nonlinear effects($\chi\neq 0$)}.\  We now examine the possible presence of magnetic energy self-trapping of an initially localized excitation. 
By means of an adequate antenna, we create an initial magnetic localized excitation by inducing a single current around one of the rings (t $n=n_{0}$).
The initial conditions are $q_{0}(0)=0, \dot{q}_{0}(0)\neq 0$. We are interested in monitoring the amount of magnetic energy that remains at the initial ring for a long time. To estimate the trapping of magnetic energy, we use the long-time average of the fraction of magnetic energy residing at the initial site:
\be 
\langle U_{0} \rangle = {1\over{T}} \int_{0}^{T} (h_{0}/H_{0})\ dt,\label{U0}
\ee
where,
\begin{eqnarray}
h_{0}&=&(1/2) ( q_{0}^2+(1+2 \lambda) \dot{q}_{0}^2 +\nonumber\\
     & & + \lambda \dot{q}_{0}\sum_{m} K(m) (\dot{q}_{m}-\dot{q}_{0}))+(\chi/4) q_{0}^4,
\end{eqnarray}
and
$H_{0}=(1/2) \dot{q}_{0}(0)^2 (1-2 \lambda -\lambda \sum_{m}K(m))$.
Numerical computations of (\ref{U0}) show that if the initial ring possesses gain(loss), $\langle U_{0} \rangle$ diverges (tends to zero) at large times. This can be explained as a direct consequence of nonlinearity.  As is well-known, for DNLS-type equations (like eq.(\ref{eq1})), nonlinearity tends to decouple the sites, favoring the self-trapping of energy, for large enough $\chi$ values. In our case, 
this means that the dynamics is, to a first approximation,  similar to that of a single SRR oscillator with gain/loss. When the initial site has gain, this site will accumulate energy that will not have enough time to get distributed to its neighbors due to the decoupling effect. This induces a steady increase in magnetic energy. On the contrary, if the initial site is lossy, it will have no time to compensate for this loss, leading to a steady energy decrease. The accumulated energy per cycle can be estimated by setting $T=2 \pi$. Results are shown in Fig. 5 which shows 
$\langle U_{0} \rangle$ as a function of $\gamma$ for several $s$ values. We see that $\langle U_{0} \rangle$ increases monotonically with the gain/loss parameter, with the fractional exponent playing no significant role.
%%%%%%%%%%%%%%%%%%%%%%%%%%%%%%%%%%%%%%
\begin{figure}[t]
 \includegraphics[scale=0.25]{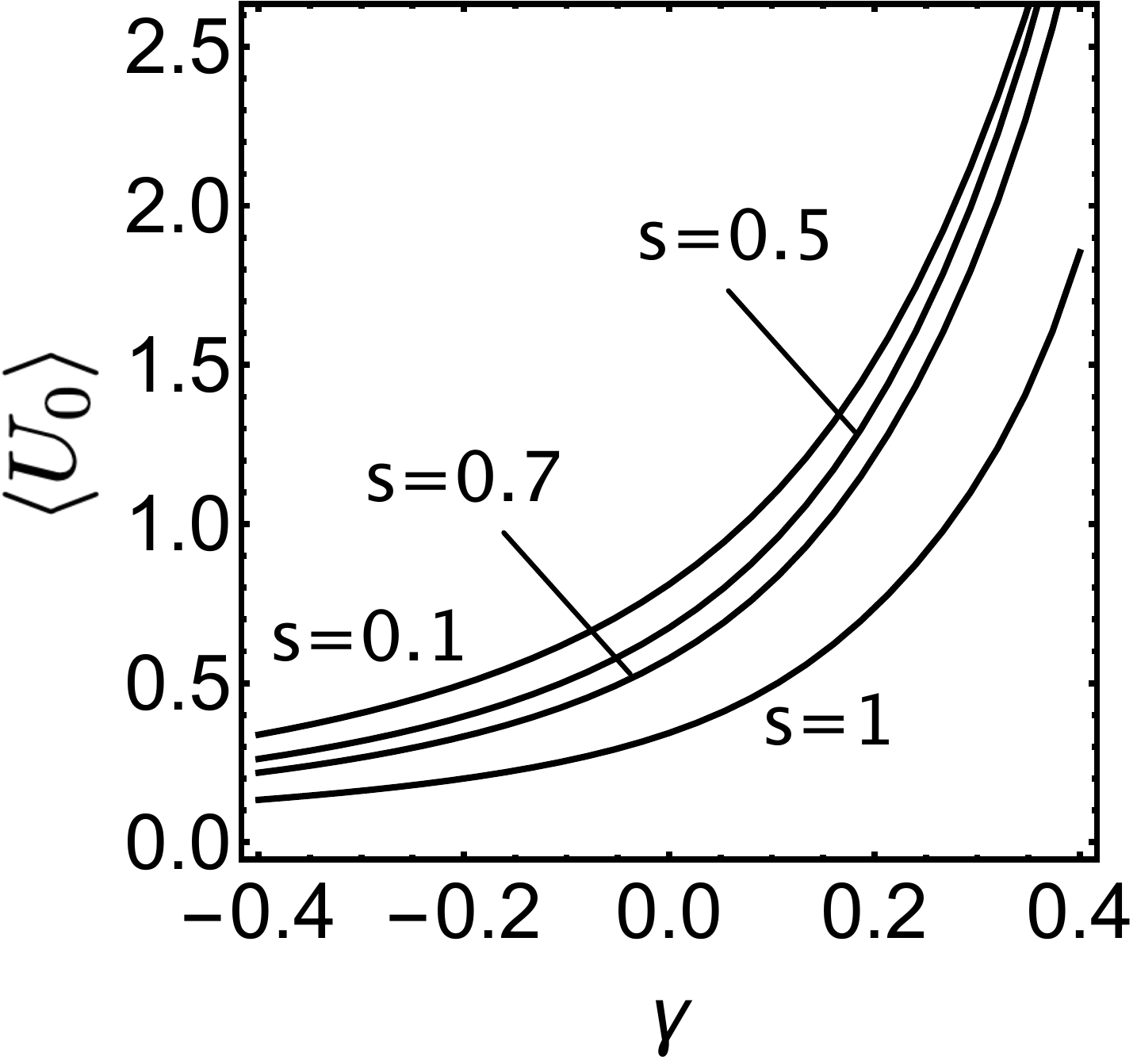}
  \caption{Average magnetic energy per cycle on initial site as a function of gain/loss, for several fractional exponent values. $(\lambda=0.2)$}
  \label{fig4}
\end{figure}
%%%%%%%%%%%%%%%%%%%%%%%%%%%%%%%%%%%%%%%%%%%%%%%%

{\em Conclusions}. We have examined the physics of a magnetic metamaterial modeled 
as an extended split-ring resonator array by using a fractional form for the discrete Laplacian operator. In the absence of nonlinearity, a closed-form expression for the dispersion of plane waves as a function of the fractional exponent is found in terms of well-known special functions. From this, we computed a stability diagram showing the regions in gain/loss-fractionality space where ${\cal PT}$ symmetry is maintained and, therefore, the system displays bounded dynamics. For $s=1$ and as soon as $\gamma$ is nonzero, the system becomes unstable for $s=1$. This agrees with previous computations for non-fractional tight-binding  arrays\cite{PTkevrekidis}. For smaller fractional exponents, the gain/loss must increase from zero in order to maintain the ${\cal PT}$ symmetry. 
The density of states is obtained in closed form and shows that the main effect of fractionality is the flattening of the two bands, while the presence of gain/loss effects alters the width of the bandgap. In the presence of nonlinearity, an examination of the magnetic energy accumulated on a given initial site as a function of time reveals that it diverges (converges to zero) when this site has gain (loss). It seems this is mainly due to nonlinearity, which influences the effective coupling between the initial site and its neighbors. For large enough nonlinearity, the dynamics can be approximated to that of a single nonlinear SRR with gain/loss, weakly coupled to its neighbors. It seems that the main effect of fractionality in this classical system is to open a window in phase space where the system maintains ${\cal PT}$ symmetry.

\vspace{0.5cm}
{\bf Acknowledgments}\\
\vspace{0.5cm}

This work was supported by FONDECYT grant 1250059 and by Vicerrectoria de Investigacion y Desarrollo (VID) de la Universidad de Chile, grant ELN10/24.
\vspace{0.5cm}

{\bf Declaration of competing interest}\\

The authors declare that they have no known competing financial interests or personal relationships that could have appeared to influence the work reported in this paper.
\vspace{0.5cm}

{\bf Data availability}\\
No data was used for the research described in this article.


\begin{thebibliography}{99}

\bibitem{fluid2}
L. A. Caffarelli, A. Vasseur, Drift diffusion equations with fractional diffusion and the quasi-geostrophic equation, Ann. of Math. {\bf 171}, 1903 (2010).

\bibitem{metzler}
R. Metzler, J. Klafter, The random walk's guide to anomalous diffusion: a fractional dynamics approach, Phys. Rep. 339 (2000) 1-77.


\bibitem{sokolov}
I. M. Sokolov, J. Klafter and A. Blumen, Fractional kinetics, Physics Today, {\bf 55}, 48 (2002).

\bibitem{zaslavsky}
G. M. Zaslavsky, Chaos, fractional kinetics, and anomalous transport, Phys. Rep. {\bf 371}, 461 (2002).

\bibitem{shlesinger}
M. F. Shlesinger, G. M. Zaslavsky, and J. Klafter, Strange kinetics, Nature {\bf 363}, 31 (1993).

\bibitem{laskin}
N. Laskin, Fractional quantum mechanics, Phys. Rev. E {\bf 62}, 3135 (2000).

\bibitem{laskin2}
N. Laskin, Phys. Rev. E {\bf 66}, 056108 (2002).

\bibitem{levy}
N. C. Petroni and M. Pusterla, Levy processes and Schr\"{o}dinger equation, Physica A {bf 388}, 824 (2009).

\bibitem{plasmas}
M. Allen, A fractional free boundary problem related to a plasma problem, https://arxiv.org/abs/1507.06289.

\bibitem{cardiac}
A. Bueno-Orovio, D. Kay, V. Grau, B. Rodriguez and K. Burrage, Fractional diffusion models of cardiac electrical propagation: role of structural heterogeneity in dispersion of repolarization, Journal of the Royal Society Interface 11(97), 20140352 (2014).

\bibitem{invasion}
H. Berestycki, J.-M. Roquejoffre, and L. Rossi, The influence of a line with fast diffusion on Fisher-KPP propagation, J. Math. Biol. {bf 66}, 743 (2013).

\bibitem{circuits}
J. Schindler, Ang Li, M.C. Zheng, F.M. Ellis, T. Kottos, Experimental study of active LRC circuits with ${\cal PT}$ symmetries, Phys. Rev. A {\bf 84} (2011) 040101(R).

\bibitem{optics1}
A. Guo, G.J. Salamo, D. Duchesne, R. Morandotti, M. Volatier-Ravat, V. Aimez, G.A. Siviloglou, D.N. Christodoulides, Observation of ${\cal PT}$-symmetry breaking in complex optical potentials, Phys. Rev. Lett. {\bf 103} (2009) 093902.

\bibitem{optics2}
R. El-Ganainy, K.G. Makris, D.N. Christodoulides, Z.H. Musslimani, Theory of
coupled optical ${\cal PT}$-symmetric structures, Opt. Lett. {\bf 32} (2007) 2632.

\bibitem{optics3}
Z.H. Musslimani, K.G. Makris, R. El-Ganainy, D.N. Christodoulides, Optical soli-
tons in ${\cal PT}$ periodic potentials, Phys. Rev. Lett. {\bf 100} (2008) 030402.

\bibitem{optics4}
K.G. Makris, R. El-Ganainy, D.N. Christodoulides, Z.H. Musslimani, Beam dynam-
ics in PT-symmetric optical lattices, Phys. Rev. Lett. {\bf 100} (2008) 103904.

\bibitem{optics5}C.E. Ruter, K.G. Makris, R. El-Ganainy, D.N. Christodoulides, M. Segev, D. Kip, Observation of parity-time symmetry in optics, Nat. Phys. {\bf 6} (2010) 192.

\bibitem{MM}
N. Lazarides, G.P. Tsironis, Gain-driven discrete breathers in ${\cal PT}$-symmetric
nonlinear metamaterials, Phys. Rev. Lett. {\bf 110} (2013) 053901.

\bibitem{solid1}
N. Hatano, D.R. Nelson, Localization transitions in non-Hermitian quantum mechanics, Phys. Rev. Lett. {\bf 77} (1996) 570.

\bibitem{solid2}
Y.N. Joglekar, D. Scott, M. Babbey, Avadh Saxena, Robust and fragile-symmetric phases in a tight-binding chain, Phys. Rev. A {\bf 82} (2010) 030103.

\bibitem{SRR1}
T. J. Yen, W. J. Padilla, N. Fang, D. C. Vier, D. R. Smith, J. B. Pendry, D. N. Basov, and X. Zhang, Terahertz magnetic response from artificial materials, Science
{\bf 303}, 1494 (2004).

\bibitem{SRR2}
N. Katsarakis, G. Constantinidis, A. Kostopoulos, R. S.
Penciu, T, F, Gundogdu, M. Kafesaki, E. N. Economou, Th. Koschny and C. M. Soukoulis, Magnetic response of split-ring resonators in the far-infrared frequency regime, Opt. Lett. {\bf 30}, 1348 (2005).

\bibitem{SRR3}
M. I. Molina, N. Lazarides, and G. P. Tsironis, Bulk and surface magneto-inductive breathers in binary metamaterials, Phys. Rev. E {\bf 80}, 046605 (2009).

\bibitem{negative_refraction}
R. A. Shelby, D. R. Smith, S. Schultz, Experimental verification of a negative index of refraction, Science {\bf 292}, 77 (2001).

\bibitem{losses1}
L. Esaki, New Phenomenon in Narrow Germanium p - n Junctions, Phys. Rev. {\bf 109}, 603 (1958).

\bibitem{losses2}
T. Jiang, K. Chang, L.-M. Si, L. Ran, and H. Xin, Active Microwave Negative-Index Metamaterial Transmission Line with Gain, Phys.
Rev. Lett. {\bf 107}, 205503 (2011).

\bibitem{shadrivov}
A. A. Zharov, I. V. Shadrivov, and Y. S. Kivshar, Nonlinear Properties of Left-Handed Metamaterials, Phys. Rev. Lett. {\bf 91}, 037401 (2003).

\bibitem{gts_2006}
N. Lazarides, M. Eleftheriou, and G. P. Tsironis, Discrete Breathers in Nonlinear Magnetic Metamaterials, Phys. Rev. Lett. {\bf 97}, 157406 (2006).

\bibitem{discrete laplacian}
Oscar Ciaurri, Luz Roncal, Pablo Raul Stinga, Jose L. Torrea, Juan Luis Varona,
Nonlocal discrete diffusion equations and the fractional discrete Laplacian, regularity, and applications, Advances in Mathematics {\bf 330} (2018) 688.

\bibitem{PTkevrekidis}
D.E. Pelinovsky, P.G. Kevrekidis, D.J. Frantzeskakis, PT-symmetric lattices with spatially extended gain/loss are generically unstable, Europhys. Lett. {\bf 101}, 11002 (2013). 

\end{thebibliography}
\end{document}